\RequirePackage{fix-cm}
\documentclass[smallextended]{svjour3}       % onecolumn (second format)
\smartqed  % flush right qed marks, e.g. at end of proof
\usepackage{graphicx}
\usepackage{amssymb}
\usepackage{textcomp}
\usepackage{gensymb}
\usepackage{amsmath}
\usepackage{xcolor}
\usepackage{multirow}
\usepackage{lineno}
\usepackage{url}
\usepackage{natbib}

\newcommand{\Prob}{\mathbb{P}}

\begin{document}

\title{Openness and Reproducibility: Insights from a Model-Centric Approach
%ALTERNATIVE TITLE SUGGESTION: Are Exact Replications and Complete Openness Necessary Conditions for a Reproducible Scientific results?
%\thanks{(Thanks omitted for review.)
%Research reported in this publication was supported by the National Institute Of General Medical Sciences of the National Institutes of Health under Award Number P20GM104420. The content is solely the responsibility of the authors and does not necessarily represent the official views of the National Institutes of Health.}
}


\author{%
        %Anonymous for Review
        Bert Baumgaertner \and
		Berna Devezer \and 
		Erkan O. Buzbas \and 
		Luis G. Nardin %
}


\institute{
%Anonymous for Review
Bert Baumgaertner \at Department of Politics and Philosophy, University of Idaho, Moscow, ID 84844-1104 USA\\\email{bbaum@uidaho.edu}           
          \and
        Berna Devezer \at Department of Business, University of Idaho, Moscow, ID 84844-1104 USA\\ \email{bdevezer@uidaho.edu}
          \and
          Erkan O. Buzbas \at Department of Statistical Science, University of Idaho, Moscow, ID 84844-1104 USA\\ \email{erkanb@uidaho.edu}	       \and
		   Luis G. Nardin \at Computer Science, National College of Ireland, Dublin, Ireland\\ \email{gnardin@gmail.com} 
}

\date{Received: date / Accepted: date}

\maketitle

\begin{abstract}
This paper investigates the conceptual relationship between openness and reproducibility using a model-centric approach, heavily informed by probability theory and statistics. We first clarify the concepts of reliability, auditability, replicability, and reproducibility--each of which denotes a potential scientific objective. Then we advance a conceptual analysis to delineate the relationship between open scientific practices and these objectives. Using the notion of an idealized experiment, we identify which components of an experiment need to be reported and which need to be repeated to achieve the relevant objective. The model-centric framework we propose aims to contribute precision and clarity to the discussions surrounding the so-called reproducibility crisis. 


\end{abstract}

%\linenumbers


%\par
\begin{quote}
	\footnotesize{``I also know this, that, although I might teach only what is true, you must deny me faith in my words. So in order that I do not perorate, but not leaving with any faith in all my words, I transfer the authority to anyone who wishes, to come and show me if what I say concerning the findings of anatomies is really true. For I have already shown thousands of times the twin [organs] that intercede the spermatic cords from the outer horns to the inside of the uterus, be the animal a goat or an ox or a donkey or a horse, that either wooden sticks round and long, even three or four times thicker, or the ones called sword-like probes, I have positioned through the horns. And this must be shown by anyone [that follows the same experimental method] after I and my pupils have died.'' \textit{Galen of Pergamon (c.130--210 AD)}}
\end{quote}

\section{Introduction}

In the last decade, scientists have discussed whether science is facing a reproducibility crisis~\citep{Baker2016} and numerous explanations have been given for the proliferation of irreproducible results~\citep{Heesen2018,Munafo2017,Spellman2015}. Many of these 
explanations suggest that irreproducibility is the consequence of methodological and cultural practices that are erroneous at individual or system levels. %one or more errors that cause divergence from ideal science. 
Well-known examples of such questionable research practices include HARKing, p-hacking, and publication bias~\citep{Bishop2019,Spellman2015}. HARKing (hypothesizing after results are known) involves presenting a post hoc hypothesis conditional on observing the data as an a priori hypothesis~\citep{Kerr1998,Munafo2017}. P-hacking is a form of data dredging to find statistically significant results and is a misapplication of proper statistical methodology~\citep{Bruns2016,Munafo2017}. Publication bias involves omitting studies with statistically nonsignificant results from publications and is primarily attributed to flawed incentive structures in scientific publishing~\citep{Munafo2017,Open2015}. The common denominator across these three phenomena is a lack of transparency (of hypotheses, analyses, and studies, respectively) in research reporting. 

As such, openness is sometimes touted as a remedy to the supposed crisis of irreproducibility~\citep{Collins2014,Iqbal2016,Nosek2015}. For example, the National Academies of Sciences, Engineering, and Medicine view data sharing as a prerequisite for reproducible results~\citep{national2017fostering}. A number of tools are becoming increasingly available to make numerous aspects of science open~\citep{Ioannidis2014,Munafo2017,Nosek2015,Nosek2018}. As the thinking goes, the building blocks of a reproducible science are replication studies~\citep{Earp2015,Lebel2011,Schmidt2009}, and  transparency makes replication studies possible by making research materials and processes explicit and accessible. The link between transparency, replication studies, and reproducibility is not straightforward, however. For example, results can be reproduced from experiments that are not completely transparent (say, because the same materials in question are not accessible, or methods are unknown) and reproducibility can be difficult to achieve, even when experiments are completely transparent and exact replication studies can be conducted. 

We are not convinced by the picture that openness, which would correct questionable research practices, is the straightforward solution to the purported reproducibility crisis. Even if openness were the solution, it is not clear to us why it would be.  Our problem is an intellectual one: if we lack a clear understanding of the relationship between transparency, replications, and reproducibility, it is not clear how or why openness is supposed to help. Before calling foul on scientists or scientific practice, it behooves us to better understand these relationships and set our expectations right for a reproducible science. That is the aim of this paper.

To set the stage, we first present a toy example and derive a number of relatively benign observations and conclusions regarding replications and reproducibility, drawing from probability theory and well-known facts in statistics (Section~\ref{toy}). This motivates our conceptualization of an idealized experiment and definitions of reliability, auditability, replicability, and reproducibility (Section~\ref{Concepts}). We are then in a position to present our analysis of openness, which comes in two parts. The first part emphasizes the epistemic aspect of reproducibility, which illuminates a historical debate between Newton, Goethe, and others (Section~\ref{open}). The second part provides a more detailed account of the components of an experiment that need to be shared (or don't) given some objective (Section~\ref{components}).

Our analysis makes little to no mention of hypotheses. This is not an oversight. Our analysis is situated in what we call a model-centric view of science. From this perspective, scientific progress is made when old models are replaced by new ones and experiments are performed in order to compare models. The idea that scientific meaning is carried by models or that models play a central role otherwise is not novel and has been discussed both in scientific and philosophical literature~\citep{Glass2008,Taper2011,weisberg2012simulation}. This approach is distinct from a hypothesis-centric approach and we contrast these two in the discussion section (Section~\ref{discussion}).

%--------------------
\section{Stage setting: A toy example and initial observations}
\label{toy}
Consider a large population of ravens where each raven is either black or white. Our goal is to estimate the true proportion of black ravens in the population, denoted by $\pi,$ given a random sample of ravens from this population. We assume that there is no classification or measurement error. That is, each observed raven can be identified correctly as black or white. We also assume that the population is well-mixed and observations are independent, each with probability of being black equal to the true proportion of black ravens in the population. 

This initial setup is generalizable and quite mundane from a statistical perspective. We could have considered any phenomenon that can be observed repeatedly, where a decision has to be made under uncertainty using observations. The probability calculus quantifies uncertainty, and statistical methods based on probabilistic models provide the machinery to perform inference about aspects of the mechanism generating these observations. Examples of these aspects include estimating a parameter or predicting future observations under an assumed probability model, or selecting between competing probability models. 

With this in mind, we now consider the following two experiments that could be conducted within the paradigm of our toy example.

\begin{enumerate}
	\item[]{\textbf{Experiment 1.}} A simple random sample of $n$ ravens is collected. We observe $b$ black ravens in the sample. The likelihood of the observed data conditional on $\pi$ and $n$ is given by the binomial probability model, and is equal to
	\begin{equation}\label{eq:likbin}
	\Prob(b|\pi,n)=\binom{n}{b}\pi^b(1-\pi)^{n-b}.
	\end{equation}
	\item[]\textbf{Experiment 2.} A simple random sample of ravens is collected until $w$ white ravens are observed. We observe $b$ black ravens such that $b+w=n.$ The likelihood of the observed data conditional on $\pi$ and $w$ is given by the negative binomial probability model, and is equal to
	\begin{equation}\label{eq:liknegbin}
	\Prob(b|\pi,w)=\binom{n-1}{w-1}\pi^b(1-\pi)^w.
	\end{equation}
\end{enumerate}
%
%
%-------------------------------------------------

This setup is not novel -- it could be found in a standard statistics text. Nevertheless, several observations about the setup are important to make explicit, as they provide us with conclusions that help guide our thinking about reproducibility and openness.

\begin{enumerate}
	\item[] {\bf Observation 1:} The probability models in Equations~\eqref{eq:likbin} and \eqref{eq:liknegbin} are different because the parameter vectors $(\pi,n)$ and $(\pi,w)$ are different in these two models. Yet the population proportion of black ravens $\pi$ is a parameter of both models and therefore can be estimated. Assume we estimate the proportion of black ravens by the well-known statistical method of the maximum likelihood (ML). The ML estimate $\hat{\pi}_{ML}$ is the proportion of black ravens that maximizes the likelihood of the observed data. The estimates for $\pi$ under model \eqref{eq:likbin} and model \eqref{eq:liknegbin} are both equal to $b/n.$  
	\item[]{\bf Conclusion 1:} Some results of Experiment 1 can be reproduced by Experiment 2 even if the models in these experiments are different. In other words, the model in Experiment 2 is not required to be the identical model as in the Experiment 1 to reproduce some of its results.  
	\item[] {\bf Observation 2:} Conclusion 1 cannot solely be explained by the use of the identical method in both experiments because ML is applied on realizations of different random variables in two models. To see this, note that the maximum number of black ravens that can be observed in Experiment 1 is $n$ but in Experiment 2 it is the maximum number of black ravens in the population. If we use $\hat{\pi}_{ML}$ in Experiment 1 and $\hat{\pi}_{MM}$ in Experiment 2, where MM denotes the method of moments estimator, the estimates are still both equal to $b/n$, even though MM is based on a different principle than ML. 	\item[] {\bf Conclusion 2:}  
	The method in Experiment 2 is not required to be the identical method as in Experiment 1 to reproduce some of its results.   
	\item[] {\bf Observation 3:} The stopping rule of data collection in Experiment 1 is different than the stopping rule in Experiment 2. Experiment 1 stops when $n$ ravens are observed. Experiment 2 stops when $w$ white ravens are observed. Thus, the data structures in Experiment 1 and Experiment 2 are different. 
	\item[] {\bf Conclusion 3:} The data structure in Experiment 2 is not required to be identical to the data structure in Experiment 1 to reproduce some of its results. 
	\item[] {\bf Observation 4:} To assess whether the estimate $b/n$ obtained in Experiment 1 is reproduced in Experiment 2, Experiment 2 must know the result of Experiment 1.   
	\item[] {\bf Conclusion 4:} Experiment 2 needs sufficient background information about results of Experiment 1 to assess whether the results are reproduced. Thus, the background information used in Experiment 1 and Experiment 2 must be different from each other.
	\item[] {\bf Observation 5:} Assume we repeat Experiment 1 a large number of times and we choose the estimator of $\pi$ as: 
	$$\hat{\pi}=\frac{b}{n}+\mathbf{I}_{\{1^{st}\;raven\;is\;black\}},$$

	where $\mathbf{I}_{\{A\}}=1$ if $A,$ and else zero. This estimator is equal to its true value $\pi$ in $100(1-\pi)\%$ on average and equal to $2\pi$ in $100\pi\%$ of the experiments on average.
	\item[]{\bf Conclusion 5:} True results are not always reproducible. For example, if the true population of black ravens is 0.8 and our experiments return a smattering of values close to the truth but only one that is exactly 0.8, then only one result is true. Note that ``true'' here involves the idea of accuracy because the estimates are real numbers. Providing a range of values instead would sacrifice accuracy to ensure truth: in the extreme, if we say that the proportion of black ravens is between 0 and 1, we guarantee a true claim, but it is not informative. For ease of exposition, we continue to use ``true'' with the precision provided by real numbers and handle the idea of closeness in other ways.
	\item[] {\bf Observation 6:} Assume we repeat Experiment 1 a large number of times and we choose the estimator of $\pi$ as: 
	$$\hat{\pi}=\frac{1}{2}.$$
   There is no statistical reason for this estimator to be equal to its true value $\pi$ since it is not using the observations as a representative sample from the population. However, we will always return the same value if we fix the sample size. 
	\item[]{\bf Conclusion 6:} Perfectly reproducible results may not be true. Hence reproducibility is not sufficient for truth. 
\end{enumerate}

A few remarks regarding the simplicity of our toy example. Observation 5 presents a biased estimator that only sometimes equals the true parameter value. While this may or may not be a realistic choice of an estimator, our point in this observation is that the observed rate of reproducibility is a function of the methods that we apply on data to make inference and the true rate of reproducibility. Theoretically speaking, we cannot always expect true results to be reproducible. Similarly, sampling error and model misspecification~\citep{Box1976,Dennis2019} present other potential reasons for why true results may not always be reproducible. Observation 6, on the other hand, is in tension with actual scientific practice, which uses observations as a way for models to make ``contact'' with the world. Though contrived, this conclusion, as well as the others, carry over to relevant aspects of real science. For example, the estimator in observation 6 could be a complicated algorithm with an unintentional ``error'' in the code implementation could have the same effect. It is thus possible that two separate labs using the same code reproduce one another's results, not because they are accessing the truth, but because the same error is biasing the process to provide the same result. Of course code verification and validation processes will help minimize the instantiation of such possibilities. This acknowledges the point. Reproducibility may occur because of aspects of experiments or analyses and not ``nature'' (or whatever scientists are trying to make contact with)\footnote{Ian Hacking has referred to this phenomenon as ``the self-vindication of the laboratory sciences''~\citep{Hacking1992}.}.

In brief, it is possible to reproduce the results of an experiment in a variety of ways, even in experiments that are not carbon copies of the original experiment. In order to provide a more detailed analysis of what does and doesn't need to be copied, and relatedly made open, we first provide some additional clarification of the idea of an experiment and define some key terms.

\section{Concepts and Terminology}
\label{Concepts}
We start with the notion of idealized experiment, which is central to many forms of scientific inquiry~\citep{Devezer2018}. Given some background knowledge $K$ on a natural phenomenon, a scientific theory makes a prediction, which is in principle testable using observables, the data $D$. A mechanism generating $D$ is formulated under uncertainty. This mechanism is represented as a probability model $M_\theta$ parametrized by $\theta.$ The extent to which, parts of $M_\theta$ are relevant to the prediction are confirmed by $D$ is assessed by a fixed and known collection of methods $S$ evaluated at $D.$  We denote $(M_\theta,D,S,K)$ by $\xi,$ an idealized experiment\footnote{Our conceptualization of the idealized experiment follows a parallel to~\citet{Hacking1992}'s taxonomy of laboratory experiments. Our $K$ and $M_\theta$ would be subsumed under his \textit{ideas}, $S$ under \textit{things}, and $D$ under \textit{marks}.}.

\begin{figure}
	\begin{center}
		\includegraphics[width=\textwidth]{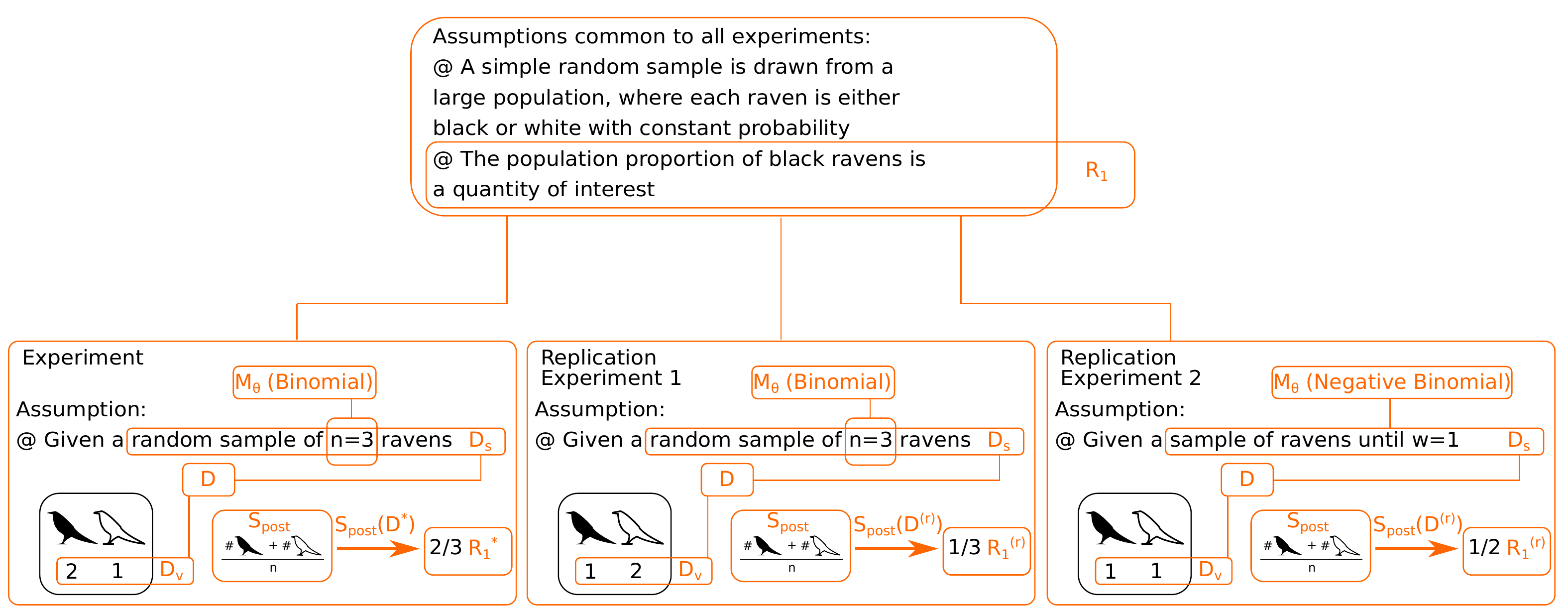}
		\caption{Elements of three idealized experiments: experiment, replication experiment 1, and an alternative replication experiment 2.}
		\label{fig:elements}
	\end{center}
\end{figure}

We further refine two components of $\xi.$ First, we let $D\equiv\{D_v,D_s\}$ where $D_v$ denotes the observed values, and $D_s$ denotes the structural aspects of the data, such as the sample size, number of variables, units of measurement for each variable, and metadata. Second, we let $S\equiv\{S_{pre},S_{post}\}$ where $S_{pre}$ denotes the procedures, instruments, experimental design, and tools used prior to and necessary for obtaining $D_v$, and $S_{post}$ denotes the analytical tools and procedures applied to $D$ once it is obtained. We define $R_i$ as a result which is obtained by applying $S_{post}$ to $D.$ We denote the set of all results obtainable from an experiment as $R\equiv\{R_1, R_2, R_3,\cdots\}.$ Figure~\ref{fig:elements} shows these elements in the context of our toy example of Section \ref{toy}.

%-------------------------------------------------
%-------------------------------------------------

Using this notion of an idealized experiment, we adopt the following definitions. We will clarify why in some cases we define a term differently from in relevant literature. 

\begin{itemize}
	
	\item [] \textit{Reliability}: Propensity of a method $S$ to produce consistent results given the same inputs or initial conditions. Conditional on the same observation in the sample space, $M_\theta,$ and $K,$ a method $S_{pre}$ is reliable if it consistently produces $D$. A method $S_{post}$ is reliable if applying $S_{post}$ to $D$ consistently yields $R.$ For the rest of this manuscript, we make the simplifying assumption that $S$ is sufficiently reliable.
	
	\item [] \textit{Auditability}: The accessibility of all necessary information regarding the components of $\xi$ so that $S_{post}$ can be applied to $D$ to obtain $R$ independently of $\xi.$ Auditing is a procedure of screening for certain errors, including human and instrumental, that may be introduced in the process of obtaining $R.$ Examples are data entry and programming errors. If $S_{post}$ is not reliable, auditability of $\xi$ will not be affected but the auditing process will also be less reliable because it may not consistently yield $R$. 
	
	\item [] \textit{Replicability}: %An experiment $\xi'$ defined by $(M'(\theta'),D',S',K')$ that aims to reproduce $R_i$ is called a replication experiment. 
	An experiment $\xi$ is replicable if information about the necessary components of $\xi$ to obtain some $R_i$ is available and if these components can be duplicated, copied, or matched in an independent experiment $\xi'.$ A replication experiment $\xi'$, generates $D'$ independent from $D,$ conditional on the true data generating mechanism. We use $R_i$ instead of $R$ in this definition because $\xi'$ might only be interested in a subset of the results of $\xi.$ A common interpretation of $\xi'$ would be $(M_\theta,D',S,K)$, where the replication experiment differs from $\xi$ only in data values $D'_v$ while duplicating all other components of $\xi.$ Our analysis in section~\ref{components} diverges from this view of replication studies and brings a more fine-tuned understanding of which components of an experiment need to be duplicated or matched for replicability.
	
	\item [] \textit{Reproducibility}: The rate of $R_i$ being reproduced. We say that $R_i$ is reproduced by $R_i'$ if $M_\theta$ and $M_\theta'$ are confirmed or disconfirmed in the same direction in a probabilistic confirmation sense, such that $R_i$ and $R_i'$ are deemed equivalent. For example, if $R_i$ is an estimate of parameter $\theta$, then $D$ confirms $\theta$ if the probability of $\theta$ after observing the data, $\Prob(\theta|M_\theta,D, K)$, is greater than the probability of $\theta$ before observing the data, $\Prob(\pi|M_\theta,K).$ Here, $R_i'$ reproduces $R_i$ if $\Prob(\theta|M_\theta',D', K')>\Prob(\theta|M_\theta',K').$ %Reproducibility is often differentiated from robustness, where the later suggests that the results can be confirmed even with some deviation of the original experiment (e.g., different types of instruments, statistical methods, or samples from different populations). We will sometimes use the term `repeatability' interchangeably with `reproducibility'.
	In order to reproduce $R_i$ for the right reasons, $S$ must be sufficiently reliable. 
\end{itemize}

Notice that our definition of reproducibility focuses on the end products of experiments, the results, and not the other components of experiments that bring those products about. This choice is fitting given the etymology of the term ``reproduce''. Our choice for our definition of replicability similarly respects its etymology---it incorporates the notion to repeat something, in our case it is the components of an experiment. While our use of these terms is consistent with, if more refined than, other work~\citep{Leonelli2018,Radder1992,Radder1996,Open2015}, there is considerable variation in how these terms are used in the scientific literature~\citep{sep-scientific-reproducibility,Penders2019,Stodden2011}. We aim to sidestep potential confusion by laying out the definitions as we have and adhering to them for the remainder of this paper.
	
 From our definitions, we conclude that auditability is not necessary for replicability or reproducibility. For example, to audit $R_i$, we need to examine $D$ and implement $S_{post}$ on $D$. To replicate $\xi$ or to reproduce $R_i$, on the other hand, we do not need to know $D.$ Moreover, auditability is not sufficient for reproducibility either. A replication experiment $\xi'$ includes new data $D'$ that is generated by the true data generating mechanism. Even if $\xi$ is auditable, $R_i$ may not be reproduced by $R_i'$---an example of which was shown in Observation 5 in Section~\ref{toy}.
 
 Our definition of auditability closely tracks the idea of openness in science. However, whereas we have just stated that audibility is not necessary for reproducibility, the science reform movement, as described in the introduction, leads us to believe that openness is necessary for reproducibility. 
 %Replication experiments that fail to reproduce results are often taken to be sufficient to cast doubt on earlier results. Moreover, failing to share data and code (that performs the analysis supporting the conclusions of a study) is often viewed as a potential contributor to the problem of reproducibility, by making ``replication and validation of results difficult or impossible''~\citep[p.~87]{national2017fostering}. (THIS PAR SEEMS TO REPEAT WHAT WE HAVE IN THE INTRO. DO WE NEED IT HERE?)
 The tension we have built between auditability, openness, and reproducibility provides an opportunity to clarify their relationship. Doing so will ultimately lead us to a better understanding of reproducibility and the putative crisis. In the next section we work through a thought experiment to illustrate an important epistemic aspect of reproducibility that helps us enrich the concept of openness.

%In this paper, we provide precise technical definitions for the concepts that we make use of and build a system using them. 
%This precision provides us a solid ground and accurate results about scientific inquiry and reproducibility within the assumptions of our system.     

%-----------------------------
\section{Open Science as a Logical Necessity to Epistemic Reproducibility} 
\label{open}

We expand our toy example of ravens with a thought experiment, ``the reproducibility collaboratorium'' to distinguish  between two types of reproducibility: {\em in-principle} and {\em epistemic}. We argue that {\em open science}, which makes the necessary components\footnote{We further explain what we mean by ``necessary components'' in Section~\ref{components}.} of an experiment available for use by others, is a logical necessity for {\em epistemic} reproducibility of research results.

We consider two collaboratoriums, a {\em closed collaboratorium} and an {\em open collaboratorium} and imagine the following scenario common to both collaboratoriums: Each collaboratorium consists of Lab 1 and Lab 2 that conduct Experiment 1 and Experiment $1'$, respectively. Experiment $1'$ is conducted after Experiment 1, and ravens are sampled from one large population. All four labs assume identical models and data structure, and employ identical methods with the goal of estimating the population proportion of black ravens using their observations. Further, we assume that the number of black ravens observed in all four labs is the same. 
%(THIS WAS UNINTENTIONALLY A LITTLE MISLEADING. WE'RE NOT TALKING ABOUT THE EXPERIMENTS 1 AND 2 STATED IN THE TOY EXAMPLE BUT IT MADE THE READER THINK SO. SO I CHANGED IT TO EXPERIMENTS 1 AND 1$'$ SINCE WE EXPLAINED WE USED PRIME TO DENOTE A REPLICATION EXPERIMENT BEFORE.)

In the closed collaboratorium, Lab 1 and Lab 2 are isolated from each other and there is no information flow from Lab 1 to Lab 2.  Crucially, because of this lack of information flow, Experiment $1'$ will match all the elements of Experiment 1 that are relevant to estimating the proportion of black ravens in the population by {\em chance}. 
Such a match is improbable since there are many reasonable ways of conducting an experiment. Nevertheless, since Experiment 1 and Experiment $1'$ use identical models and methods and observed the same number of black ravens in the sample, they return the same estimate of the population proportion of black ravens. However, Experiment $1'$ does not have any information pertaining to the results of Experiment 1, and thus Lab 2 is in a position neither to learn from the results of Experiment 1, nor to claim that it reproduced the result of Experiment 1. If an external observer were to observe the experiments conducted in both labs, they could first learn from the result of Experiment 1. Starting with an updated view about the proportion of black ravens provided by the result of Experiment 1, they could then use the number of ravens observed in Experiment $1'$ to conclude that the result of Experiment 1 is indeed reproduced by Experiment $1'$.  When there is no information exchange between the labs, however, there is no meaningful {\em epistemic} interaction between Experiment 1 and Experiment $1'$. 

This closed collaboratorium example highlights two important points: (1) If there is no open science in the sense of information flow from one experiment (or lab) to the next, it is improbable (but still possible) for a replication experiment to take place, and (2) the result of Experiment 1 can be said to be reproduced by Experiment $1'$ only if the result of Experiment 1 is available to Experiment $1'$. In order to acknowledge these points, we say that a result can only be \textit{in-principle} reproducible if there is no epistemic exchange between the labs which would accumulate evidence, with the exception of via some omniscient external observer.

\begin{figure}
	\begin{center}
		\includegraphics[width=\textwidth]{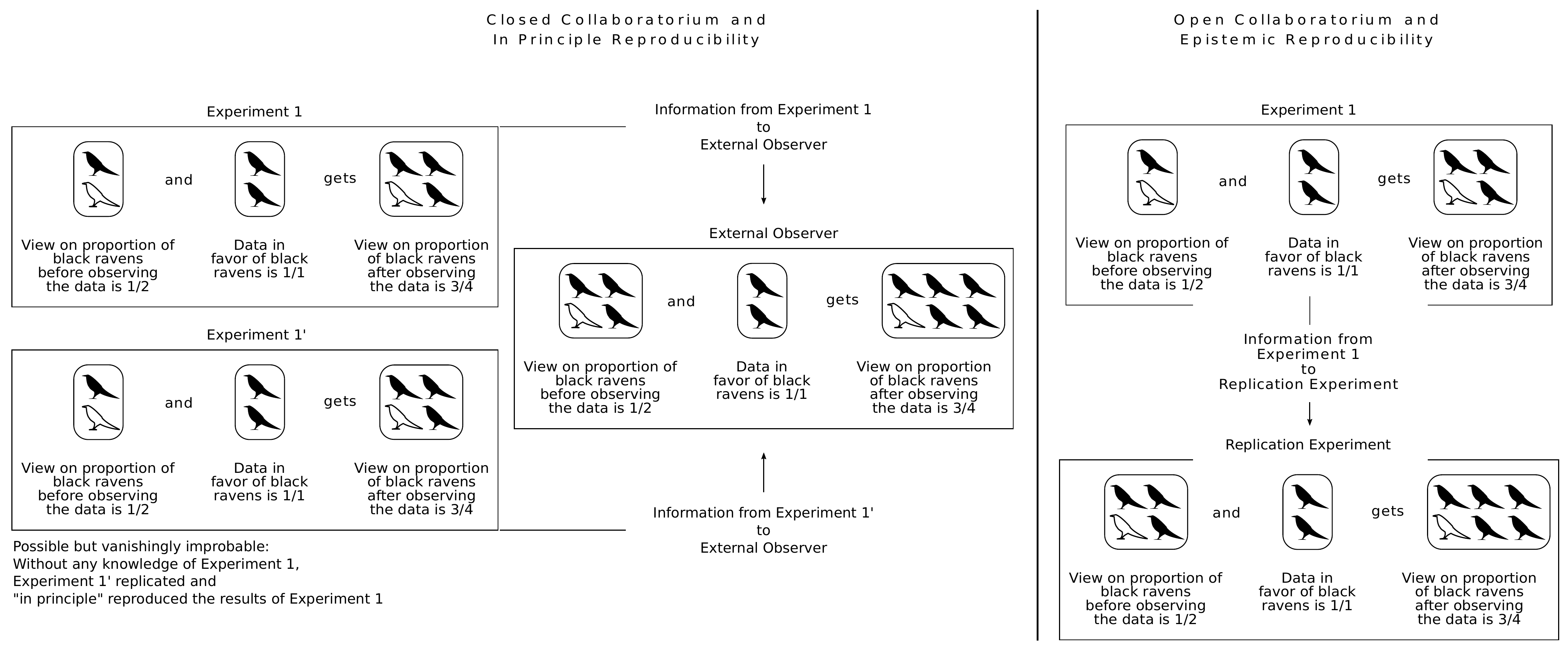}
		\caption{Closed collaboratorium: Experiment 1 starts with prior view of $1/2$ on proportion of black ravens. Observes $2$ black ravens and updates their view to $3/4.$ Identical view, model, and methods are assumed and the same data values are observed in Experiment $1'$, but in the absence of an external observer the two results cannot be connected, thus reproducibility is only {\em in principle} and evidence does not accumulate in the absence of an external observer privy to both experiments. Open collaboratorium: Experiment $1$ starts with prior view of $1/2$ on proportion of black ravens. Observes $2$ black ravens and updates their view to $3/4.$ Experiment $1'$--a replication experiment--is informed of the result, model, methods and observes the same data values. Starting with a view of $3/4$ they update the proportion of black ravens to $5/6.$ Thus, Experiment $1'$ learns from Experiment 1 in a planned manner. The two results can be connected and thus reproducibility is {\em epistemic.}}
		\label{fig:collaboratorium}
	\end{center}
\end{figure}

In the open collaboratorium we assume that Experiment 1 has a view on the population proportion of black ravens prior to observing ravens. By contrast to the closed version, however, in the open collaboratorium Lab 1 reports all information relevant to estimating the proportion of black ravens to Lab 2, which incorporates this information to conduct Experiment $1'$.  Thus Experiment $1'$ matches the elements of Experiment 1 by a kind of {\em social learning}. We assume this information is transmitted in the background knowledge of the experiment. Starting with an updated view about the proportion of black ravens in the population and conditional on the number of black ravens observed, Lab 2 could conclude that they have indeed reproduced the result of Experiment 1. Thus in the open collaboratorium there is an {\em epistemic} interaction between the two experiments which contributes to the progress of science through accumulation of evidence. In contrast to the closed collaboratorium, in the open collaboratorium replication experiments are not contingent on chance but can be routinely performed via social learning, which gives us the notion of \emph{epistemic} reproducibility. We illustrate the two collaboratoriums in Figure~\ref{fig:collaboratorium}.
%-------------------------------------------------
%-------------------------------------------------

The distinction between in-principle and epistemic reproducibility is relevant to understanding debates about reproducibility. Consider a historical example. Isaac Newton (1643--1727~AD) believed his \textit{experimentum crucis} did not need to be replicated because it already presented conclusive proof of his theory. When Anthony Lucas (1633--1693~AD) failed to reproduce and even negated Newton's results in his replication experiments, Newton was angry: [it was not] ``the number of experiments, but weight to be regarded; and where one will do, what need many?''~\citep{Newton1676}. When faced with further criticism from his opponents, Newton refused to discuss Lucas's (unsuccessful) replications and invited Lucas to talk about his original \textit{experimentum crucis} instead~\citep[p.~275]{Westfall1981}. 

Newton's anti-replication attitude was heavily criticized by Johann Wolfgang von Goethe (1749--1832~AD) who believed that single experiments or even several of them would not be enough to prove any theory, and that it was a major task of scientists to design and conduct a series of contiguous experiments, each derived from the preceding one \citep{Ribe1985}. Goethe's insistence on replication experiments echoes a long and far ranging history of advocates. Any coverage of this history here would be superficial and take us too far afield. Suffice it to say that this history is on the side of Goethe.

How can we make sense of Netwon's anti-replication attitude against a large backdrop of advocates represented by Goethe? If Newton's dismissive attitude towards replication is framed by in-principle reproducibility, we can understand why the replication of experiments is unnecessary. Newton must have already believed that his results were in-principle reproducible because of the underlying theory, hence, he did not deem any direct replication experiments necessary. This view would be consistent with his famous quote ``hypotheses non fingo'' (I feign no hypotheses) since he did not see his \textit{experimentum crucis} so much as testing a scientific hypothesis as demonstrating an already proven theory. This view, however, is not widely shared by scientists conducting experiments under uncertainty today. For most of empirical science today, scientists must exchange information and replicate each others' experiments in order to increase their confidence that new knowledge has been added (or, that they know that they have discovered some new fact) by way of reproducing results. It seems that Goethe and others were concerned with epistemic rather than in-principle reproducibility.

With epistemic reproducibility defined, we can take a closer look at the components $(M_\theta, D, S, K)$ defining $\xi$ and investigate what, more specifically, needs to be open for epistemic reproducibility. We turn to this topic in the next section.

\section{Which components of an experiment need to be open for epistemic reproducibility?} 
\label{components}

In recent years several tools have been developed to help facilitate openness in science~\citep{Collins2014,Munafo2017,Nosek2015,Wagenmakers2012}. In some cases the net is cast broadly, making as much information available as possible. In other cases intuition guides which components are relevant and need to be shared in the replication of an experiment. We are interested in using our model-centric framework to better understand what does and does not need to be made available and under what conditions.

The context of epistemic reproducibility is our starting point. In addition, our analysis will take into account our toy example and initial observations, as well as our thought experiment about collaboratoriums. Moreover, our analysis will be structured by our concept of an idealized experiment. That is, we confine ourselves to identify the components of experiments as we have conceived of them, leaving open the possibility that there are other ways of describing scientific processes relevant to replication and reproducibility.

For our analysis, we can understand an experiment $\xi$ given by $(M_\theta, D, S, K)$ as a function that takes $D$ generated from the true data generating model as a random input and produces a result $R$ as a random output. Thus, $\xi$ is a random transformation from the space of data to the space of results under the assumptions specified by $M_\theta$ (the model), $S$ (the methods), and $K$ (background knowledge). Assuming that a model in an experiment captures the true data generating mechanism, we can investigate which components of the model, the method, and the data are needed to be transmitted to reproduce the results of an experiment in a replication experiment $\xi'$. We encapsulate the transmission of these components from $\xi$ to $\xi'$ in $K'$. By definition, this makes $K$ and $K'$ different, a version of an observation we already made in sections~\ref{toy} and~\ref{Concepts}. 

Our results are grouped by components: i) (model) specific aspects of the model might have to be shared, but whether it does and which parts of a model depends on the objective; ii) (method) we specifically identify those aspects of $S'_{post}$ that need to be shared; iii) (data) here we distinguish between the structural aspects of data and the observed values, and what has to be open depends on whether we are doing an exact replication or a reproducibility experiment.

\subsection{What parts of model $M_\theta$ are needed for reproducibility?}
\label{model}

Statistical theory shows us that for $\xi'$ to be able to reproduce {\em all possible} results of $\xi,$ the specification of model $M_\theta$ up to the unknown quantities needs to be transmitted to the replication experiment, such that $M_\theta$ and $M_\theta'$ are identical models. If an aspect of $M_\theta$ that has an inferential value is not transmitted to $\xi'$, that inferential value is lost, and the results relevant to that inferential value cannot be reproduced. 

On the other hand, given an inferential objective to produce a specific result $R_i$, the aspects of $M_\theta$ that are irrelevant to that objective need not be transmitted to the replication experiment. This point is shown by Observation 1 and Conclusion 1 of our toy example in Section~\ref{toy}. When we consider estimating the population proportion of black ravens, the two models in our example are different from each other but have an identical parameter capturing the population proportion of black ravens and they both employ the number of black ravens observed in the sample in the same way for that particular objective. 

If $M_\theta$ is not identical to $M_\theta',$ then $\xi$ and $\xi'$ differ from each other with respect to the assumed data generating mechanism. What matters is whether these aspects affect the results of $S$ applied on $D$ for estimating the population proportion of black ravens. That is, even though the model in a replication experiment differs from the original, what matters is that the models share the relevant parameters, in this case the proportion of black ravens.

We can demonstrate this formally. Equation~\eqref{eq:likbin} and \eqref{eq:liknegbin} give the likelihood of observing $b$ black ravens in a sample of size $n$ under the binomial and negative binomial probability models respectively and the maximum likelihood estimate for the population proportion of black ravens under both models is $b/n$.  The reason for this is that binomial and negative binomial models are in the same likelihood equivalence class with respect to the objective of estimating the population proportion of black ravens: The maximum likelihood estimator can be derived by setting the expression resulting from taking the derivative of the logarithm of the likelihood function with respect to $\pi$ and solving for $\pi$. For Equation~\eqref{eq:likbin} we get 

\begin{equation}\label{eq:mlebin}
\frac{d}{d\pi}[\log \Prob(b|\pi,n)]=\frac{d}{d\pi}\left[\log{\binom{n}{b}}\right]+\frac{d}{d\pi}[b\log\pi] + \frac{d}{d\pi}[w\log(1-\pi)],
\end{equation}
and for Equation~\eqref{eq:liknegbin} we get 
\begin{equation}\label{eq:mlenegbin}
\frac{d}{d\pi}[\log \Prob(b|\pi,n)]=\frac{d}{d\pi}\left[\log {\binom{n-1}{w-1}}\right] + \frac{d}{d\pi}[b\log \pi] + \frac{d}{d\pi}[w\log (1-\pi)].
\end{equation}
The difference between these two equations is only in the first terms which are equal to zero. We get $\hat{\pi}=b/n$ as the unique solution in both models.

The first term in Equation~\eqref{eq:likbin} and Equation~\eqref{eq:liknegbin} determines the stopping rule of the experiments. In $\xi,$ we stop the experiment when $n$ ravens are observed and the last raven can be black or white. In $\xi'$ we stop the experiment when $w$ white ravens are observed and the last observation must be a white raven. This difference between stopping rules means that: 1) $S_{pre}$ is different from $S_{pre}'.$ 2) Under our choice of $S_{post}$ as the maximum likelihood estimator the stopping rules in two models are irrelevant for estimating the proportion of black ravens in the population.

% Examples where the choice of $S_{post}$ yield different $R_i$ and $R_i'$ in these two models can also be found. An example is frequentist hypothesis testing for the population proportion of black ravens.

We also need to distinguish between openness (and auditability) of $M_\theta$ from replicability of $M_\theta.$ In our binomial and negative binomial models, $M_\theta'$ is different from $M_\theta.$ However, these two models are compatible with respect to a certain inferential objective that allows for reproducing a specific $R_i,$ which is estimating the proportion of black ravens in the population. To establish this compatibility, $M_\theta$ should be open to $\xi'$ but does not need to be replicable or replicated. Specifically, to choose a negative binomial probability model in $\xi'$ to reproduce the estimate of the proportion of black ravens in the population obtained in $\xi$, we need to know that $\xi$ has used a binomial probability model, which ensures that $\xi'$ will use a probability model that has the same parameter---proportion of black ravens, with the exact same meaning in $\xi$. Without $M_\theta,$ this compatibility cannot be established.

This point is clearly illustrated in a recent article~\citep{Silberzahn2018} in which the same data $D$ was independently analyzed by twenty-nine research teams who were provided data and a research question that puts a restriction on which $R_i$s would be relevant for the purposes of the project. The teams were not, however, provided a $M_\theta,$ $S_{post},$ or $K.$ Teams ended up using a variety of models differing in their assumptions about the error variance and the number of covariates to analyze the same data set. The results differed widely with regard to reported effect sizes and hypothesis tests. So even when $D$ was open, the lack of specification with regard to $M_\theta$ yielded largely inconsistent results.

Taking stock, our ravens example is deliberately simple to help in our analysis. State of the art models are often complex objects. If the assumed model and its assumptions are complex, it might not always be clear which class of models contains others, and a matching model for $\xi$ may not even be available. Then, $M_\theta$ needs to be both auditable and replicable for reproducibility. This result is particularly important to communicate to scientists who primarily engage in routine null hypothesis significant testing procedures and may not be conventionally expected to transparently report their models.

\subsection{What parts of a method are needed for reproducibility?}
\label{method}

In this section, we focus on $S_{post}$ (the analytical methods applied to data) but leave $S_{pre}$ (experimental procedures to generate data) unspecified. Studying $S_{pre}$ is complicated because for a given model $M_\theta,$ the number of ways that an experiment can be designed is not well specified under statistical theory, and procedures and measurements to test the same research question can vary. Even in our simple ravens example, a raven can be observed for its color by an investigator using their eyes, but a blind investigator may opt for a mechanical pigment test. This experimental design issue is sometimes referred to as ``hidden moderators'' when explaining why results of replication experiments differ from original experiments~\citep{Baribault2018}. In addition, the issues surrounding measurement error has been studied extensively and measurement error might be a potential factor exacerbating irreproducibility~\citep{Loken2017,Stanley2014}. What we can say is that, at minimum, auditability of $S_{pre}$ appears to be essential for reproducibility. Once all experimental procedures, design details, and instruments are reported, their exact replicability becomes less of an issue for $\xi'$ to the degree that measurement error can be explicitly modeled in case of any deviation from $S_{pre}$.

Turning our attention to analytical methods applied to data, Observation 3 and Conclusion  3 in Section~\ref{toy} show that $S_{post}$ and $S_{post}'$ do not have to be identical. Some statistical methods are mathematically equivalent even though their motivations are different. For example, the maximum likelihood estimator and the method of moments estimator are equivalent in estimating the population proportion of black ravens in our toy example. We can demonstrate this formally. Consider the binomial model specified by Equation~\eqref{eq:likbin}. If $S_{post}$ is the maximum likelihood estimator motivated by the likelihood principle, the standard procedure to obtain it is to take the likelihood represented by Equation~\eqref{eq:mlebin}, setting the right hand side to zero, taking the derivative, and finally solving for $\pi:$
\begin{eqnarray*}
	\frac{d}{d\pi}\left[\log{\binom{n}{b}}\right]+\frac{d}{d\pi}[b\log\pi] + \frac{d}{d\pi}[w\log(1-\pi)]&=0,\\
	b/\pi - w/(1-\pi)&=0,
\end{eqnarray*}
so we have 
$$\hat{\pi}_{ML} = b/(b+w) \Rightarrow \hat{\pi}_{ML} = b/n.$$
On the other hand, if $S_{post}$ is the method of moments estimator, the motivation is to set the population mean equal to the sample mean and solve for $\pi.$ The population mean in a binomial model with sample size $n$ and the probability of observing a black raven $\pi$ is $n\pi$ and the sample mean is the number of black ravens in the sample $b.$ So the method of moments estimator is
$$n\hat{\pi}_{MM}=b\;\Rightarrow\; \hat{\pi}_{MM}=b/n,$$
equivalent to $\hat{\pi}_{ML}.$ 

Furthermore, other methods are mathematically equivalent even though their very interpretation of probability differs. For example, maximum likelihood estimates and the posterior mode in Bayesian inference under uniform prior distribution of parameters are equivalent regardless of the true data generating model. Conversely, there are also methods designed for a specific goal but that do not produce identical $R$ when applied to the same $D$. For example, ~\citet{Devezer2018} shows that the choice between the Akaike's Information Criterion and Schwarz criterion might influence the reproducibility of results in a model comparison.

In addition, a statistical method used to draw inferences about is often conditional on a fully specified statistical model up to a finite number of unknown parameters of that model. Consider the following situation. A scientist who designs $\xi'$ is given only the following information about $\xi:$ ``A population has only black and white ravens, and ravens are sampled to perform inference about the error on the population proportion of black ravens.'' This is an underspecified model that might easily lead to different methods of choice for $S_{post}$ and $S_{post}'$---the estimator of the error of population proportion of black ravens. The scientist might assume a large population and sample the ravens with replacement to build a binomial probability model for the data generating process. On the other hand, she might also assume a small population and sample the ravens without replacement to build a hypergeometric probability model. The error about these estimates are different even though $S_{post}$ and $S_{post}'$ might be motivated by one principle, such as maximum likelihood. Further, these differences are likely to be exacerbated if the methods are motivated by different principles. 

From these examples, we infer that $S_{post}'$ either needs to be identical to $S_{post}$ or should match it with regard to desired inference (point estimation, interval estimation, hypothesis testing, prediction, model selection) and in a way to allow for reproducing a specific $R_i.$ To match, $S_{post}$ should be open or auditable to $\xi'$ but does not need to be replicable or replicated. In order to use the method of moments estimator to estimate the proportion of black ravens in a replication experiment, we need to know that $\xi$ has used a maximum likelihood estimator. This way, it can be ensured that $\xi'$ will either use the same estimator as $\xi$ or will match it.

\subsection{What parts of data are needed for reproducibility?}
\label{data}

In section \ref{Concepts} we defined $D\equiv\{D_v,D_s\}$ where $D_v$ denotes the observed values, and $D_s$ denotes the structural aspects of the data, such as the sample size, number of variables, units of measurement for each variable, and metadata.

If $D\equiv\{D_v,D_s\}$ and $D'\equiv\{D_v',D_s'\}$ are the data obtained in an experiment and a replication experiment respectively, $D'$ is often thought of as the new data of the {\em old kind} in the sense that the values $D_v'$ are independent from the values $D_v,$ but that the data structures $D_s$ and $D_s'$ are identical. Observation 4 and Conclusion 4 of our toy example consists of a counterexample to this case where the $D_s$ and $D_s'$ can be different and the result $R_i$ is still reproduced by $R_i',$ if these differences do not affect how the method $S$ evaluates $D$ and $D'$ for the inferential objective.  

While $D_s$ does not need to be exactly duplicated in $\xi'$ for reproducibility of $R_i$, the parts of it relevant to obtaining $R_i$ need to be open. Consider a situation where some ravens cannot be classified as black or white (perhaps due to $S_{pre}$ not being sufficiently sensitive) and are recorded as missing data. In this case the number of missing data points is carried in $D_s.$ If the estimate of population proportion of black ravens is reported as $b/n,$ without $D_s,$ $\xi'$ would not know whether the number of missing data points are treated as part of $n$ or left out.  Therefore, $\xi'$ cannot ensure whether they reproduced $R_i.$  From this example, we infer that $D_s'$ either needs to be identical to $D_s$ or should match it with regard to desired inference and in a way to allow for reproducing a specific $R_i.$ For similar reasons, $D_s$ also needs to be open for the purposes of auditability of $R.$ 

Data sharing is often viewed as a prerequisite for a reproducible science~\citep{national2017fostering,Hardwicke2018,Molloy2011,Stodden2011}. Our analysis suggests this is potentially misconceived. Using the components of the idealized experiment and statistical theory, we have shown that to reproduce a result $R_i$ of $\xi:=(M_\theta,D,S,K)$, one needs to know aspects of $M_\theta,$ $D_s$, and $S$ relevant to obtaining $R_i.$ Moreover, a replication experiment $\xi'$ need not copy these aspects of $M_\theta,$ $D_s$, and $S$ to reproduce $R_i.$ We also show that having open access to $D_v$ has no bearing on designing and performing a replication experiment $\xi'$ or reproducibility of $R_i.$ $\xi'$ aims to reproduce the result, not the data. That said, openness of $D_v$ is necessary for auditability of $\xi$. Auditability, replicability, and reproducibility are distinct concepts and they need to be assessed separately when evaluating individual experiments. While some level of open scientific practices is necessary to obtain reproducible results, we argue that open data are not a prerequisite. 

There might be other benefits to open data, from auditability of results to enabling further research on the same data; however, the distinction we draw matters particularly in situations where there may be arguably valid concerns, such as ethics regarding data sharing~\citep{Borgman2012}. We recommend that open data be evaluated on its own merits which has been discussed extensively~\citep{Janssen2012} but not as a precursor of reproducibility.

\section{Discussion}\label{discussion}

%\subsection{Model-centric view of science}

Open practices in scientific inquiry have long been intuitively proposed as a key to solve the issues surrounding reproducibility of scientific results. However, a formal framework to validate this intuition has been missing and is needed for a clearer discussion of reproducibility. We have contributed to such a theoretical framework here. We finish with some discussion about how we see our project situated in the landscape of theories of science.

Within the last century, an important approach that has been taken towards understanding reproducibility was motivated by the work of Karl Popper, particularly The Logic of Scientific Discovery and the notion of falsifiability. \citet{Popper1959} states that ``non-reproducible single occurrences are of no significance to science'' (p. 86) and they would not be useful in refuting theories. Here Popper appears to rely on some notion of reproducibility to establish his falsifiability criterion of science. At the very least, Popper would agree that in order to refute a theory or [scientific] hypothesis, the experimental result must indeed be a falsifying counterexample. There is, however, a tension between Popper's emphasis on falsifiability and his skepticism towards confirmation. To generate confidence that a counterexample is genuine and not an error or one-off case is to ensure that the result is reproducible. To do this, the original experiment should be replicated. The problem here is that this process of replication and reproducibility is a kind of confirmation, one that establishes or increases confidence that a counterexample is genuine. Only once we establish that we have a genuine counterexample can we then proceed in Popperian-style towards the falsification of the hypothesis or theory in question. 

Popper's falsifiability view has had considerable impact on 20th century science and reproducibility has implicitly been accepted as part of scientific activity. Furthermore, Popper's view on falsifiability motivates what we call the hypothesis-centric approach to understanding reproducibility. On this approach, an experiment is performed in order to disconfirm a hypothesis, and thereby falsify a theory~\citep{Glass2008}. One example of taking the hypothesis-centric approach in the contemporary literature about reproducibility is~\citet{McElreath2015}. %They modeled population dynamics of scientific discovery and investigated how the evidential value of replication studies changed based on the levels of base rate of true hypotheses, statistical power, and false positive rate. They focused on a population of scientists testing a variety of hypotheses that has access to a tally of positive and negative published findings. For each hypothesis, this tally keeps track of the difference between the number of published positive and negative findings. The tally for a given hypothesis is assumed to be informative regarding its truth.

There is an alternative. Whereas Popper places a great deal of emphasis on the concepts of theories and hypotheses, much of current science, particularly given the rise of Bayesianism, focuses instead on models and model comparison~\citep{Burnham2003}. Since the exponential increase in computing power it has become possible to perform the necessary computations and complete analyses that were previously practically impossible. An important consequence of this change is the ability to consider, compare, and contrast different models that explain or predict data. As a result, vague scientific hypotheses can be given more rigorous specifications in terms of statistical models, which in turn provides precision in testing.

Our work here is an example of this alternative, \emph{model-centric} approach to study reproducibility. An experiment is performed in order to compare models and scientific progress is made as old models are replaced by new models. That is, whereas the hypothesis-centric approach typically assumes a model structure and tests a hypothesis, in a model-centric approach the whole model is considered and compared to other models. In this model-centric view, hypotheses are subsumed under models such that a hypothesis represents a specific statement about a model parameter.

In addition to being more general, a model-centric approach avoids certain challenges that a hypothesis-centric approach faces, in particular the underdetermination of theory by data and holism~\citep{Quine1976}. Theories or hypotheses can never be tested in isolation because they come as a bundle and with a group of background assumptions. For example, given evidence that is inconsistent with some hypothesis, one option is to reject the hypothesis and corresponding theory. Another option, however, is to place blame on the measuring instruments, the experimental procedures, or some background assumption. How does one decide which option to exercise? Scientists are generally adept at trouble-shooting issues and figuring out whether a particular instrument is malfunctioning, for example. They do this by holding fixed one set of assumptions (e.g., background theory and the reliability of the experimental setup) and checking the reliability of others (e.g., whether a particular measuring instrument is working as expected). And that is the point about holism: there is always some network of assumptions being held fixed when determining where to place the blame for negative results. Claiming to test a hypothesis in isolation is pretending that there is no network of assumptions, but there is, and consequently, a hypothesis cannot be tested in isolation. 

The model-centric approach does a better job of making background assumptions more explicit than the hypothesis-centric approach. Moreover, the model-centric approach is better suited to capture the scientific practice of model comparison, an integral part of the open, collaborative practices that have been proposed to solve issues surrounding reproducibility. Our analysis underscores the importance of transmitting model-specific information for reproducibility of results---a condition often readily satisfied in a model-centric framework. We believe that a hypothesis-centric approach is too impoverished to provide the necessary resources for a formal theory of such open practices. 

\section*{Conclusion}

We used our model-centric approach and formalization of reproducibility and related concepts, to distinguish between reliability, auditability, replicability, and reproducibility. The relationship among them is not as straightforward as it may seem and a need for a nuanced understanding is warranted. For example, a perfectly auditable experiment does not necessarily lead to reproducible results, and an experiment that does not open its data does not necessarily yield irreproducible results. Nevertheless, irreproducible results sometimes raise suspicion and discussions typically turn towards concerns regarding the transparency of research or validity of findings. These discussions, however, use heuristic analogs of the concepts of reliability, auditability, replicability, and reproducibility. Such heuristics might not hold and can lead to erroneous inferences about research findings and researchers' practices. Relatedly, we have provided some details regarding which components need to be made open relative to some objective, and which don't. For example, while necessary for auditability of experiments, data sharing is not a prerequisite for reproducible results, as suggested by NASEM, but other components of an experiment are. On the other hand, reporting model details, such as modeling assumptions, model structure, and parameters, becomes critical for improving reproducibility. Notably, even in recent recommendations for improving transparency in reporting via practices such as preregistration, models are typically left out while transparency of hypotheses, methods, and study design are emphasized~\citep{Nosek2018,Veer2016}.  

Our framework is useful in improving the accuracy of judgments made regarding replication and reproducibility. The literature on replication crisis cites several putative causes of irreproducibility, including p-hacking, HARKing, and publication bias. The social epistemology literature contributes other underlying causes such as the rush to publish results due to perverse reward structures within the publication system~\citep{Heesen2018}. Our analysis shows that neither elimination of questionable research practices nor correction of scientific reward structures would necessarily lead to reproducible results, as there are other impediments to reproducibility logically preceding lack of rigor or transparency. For example, the rate of reproducibility is a parameter of the system and therefore is a function of truth. Some level of irreproducibility will always remain as a component of the system and we can only hope to attain a level of reproducibility within the bounds of model misspecification and sampling error. And even under the assumption of an ideal version of science that is free of methodological and cultural errors, employs reliable methods, and does not operate under model misspecification, we might still not be able to make a true discovery despite having improved the rate of reproducibility. We are optimistic that our analysis improves the level of precision in discussions surrounding the drivers of epistemic reproducibility. %Only by a clear and impartial understanding of these drivers can we have an honest discussion regarding how to improve our scientific practices and what such improvements might might mean with regard to scientific progress.

\bibliographystyle{spbasic}      % basic style, author-year citations
\bibliography{repro}   % name your BibTeX data base

\begin{thebibliography}{47}
\providecommand{\natexlab}[1]{#1}
\providecommand{\url}[1]{{#1}}
\providecommand{\urlprefix}{URL }
\expandafter\ifx\csname urlstyle\endcsname\relax
  \providecommand{\doi}[1]{DOI~\discretionary{}{}{}#1}\else
  \providecommand{\doi}{DOI~\discretionary{}{}{}\begingroup
  \urlstyle{rm}\Url}\fi
\providecommand{\eprint}[2][]{\url{#2}}

\bibitem[{Baker(2016)}]{Baker2016}
Baker M (2016) 1,500 scientists lift the lid on reproducibility. Nature
  533(7604):452--454, \doi{10.1038/533452a}

\bibitem[{Baribault et~al.(2018)Baribault, Donkin, Little, Trueblood, Oravecz,
  van Ravenzwaaij, White, De~Boeck, and Vandekerckhove}]{Baribault2018}
Baribault B, Donkin C, Little DR, Trueblood JS, Oravecz Z, van Ravenzwaaij D,
  White CN, De~Boeck P, Vandekerckhove J (2018) Metastudies for robust tests of
  theory. Proceedings of the National Academy of Sciences 115(11):2607--2612,
  \doi{10.1073/pnas.1708285114}

\bibitem[{Bishop(2019)}]{Bishop2019}
Bishop D (2019) Rein in the four horsemen of irreproducibility. Nature
  568(7753):435--435

\bibitem[{Borgman(2012)}]{Borgman2012}
Borgman CL (2012) The conundrum of sharing research data. Journal of the
  American Society for Information Science and Technology 63(6):1059--1078,
  \doi{10.1002/asi.22634}

\bibitem[{Box(1976)}]{Box1976}
Box GE (1976) Science and statistics. Journal of the American Statistical
  Association 71(356):791--799

\bibitem[{Bruns and Ioannidis(2016)}]{Bruns2016}
Bruns SB, Ioannidis JPA (2016) P-curve and p-hacking in observational research.
  {PLOS} {ONE} 11(2):e0149144, \doi{10.1371/journal.pone.0149144}

\bibitem[{Burnham and Anderson(2003)}]{Burnham2003}
Burnham KP, Anderson DR (2003) Model selection and multimodel inference: a
  practical information-theoretic approach. Springer Science \& Business Media

\bibitem[{Collaboration et~al.(2015)}]{Open2015}
Collaboration OS, et~al. (2015) Estimating the reproducibility of psychological
  science. Science 349(6251):aac4716--aac4716, \doi{10.1126/science.aac4716}

\bibitem[{Collins and Tabak(2014)}]{Collins2014}
Collins FS, Tabak LA (2014) Policy: Nih plans to enhance reproducibility.
  Nature News 505(7485):612

\bibitem[{Dennis et~al.(2019)Dennis, Ponciano, Taper, and Lele}]{Dennis2019}
Dennis B, Ponciano JM, Taper ML, Lele SR (2019) Errors in statistical inference
  under model misspecification: evidence, hypothesis testing, and aic.
  Frontiers in Ecology and Evolution 7:372

\bibitem[{Devezer et~al.(2018)Devezer, Nardin, Baumgaertnet, and
  Buzbas}]{Devezer2018}
Devezer B, Nardin LG, Baumgaertnet B, Buzbas E (2018) Discovery of truth is not
  implied by reproducibility but facilitated by innovation and epistemic
  diversity in a model-centric framework. ArXiv e-prints
  \urlprefix\url{\url{https://arxiv.org/abs/1803.10118v2}}

\bibitem[{Earp and Trafimow(2015)}]{Earp2015}
Earp BD, Trafimow D (2015) Replication, falsification, and the crisis of
  confidence in social psychology. Frontiers in psychology 6:621

\bibitem[{Fidler and Wilcox(2018)}]{sep-scientific-reproducibility}
Fidler F, Wilcox J (2018) Reproducibility of scientific results. In: Zalta EN
  (ed) The Stanford Encyclopedia of Philosophy, winter 2018 edn, Metaphysics
  Research Lab, Stanford University

\bibitem[{Glass and Hall(2008)}]{Glass2008}
Glass DJ, Hall N (2008) A brief history of the hypothesis. Cell 134(3):378--381

\bibitem[{Hacking(1992)}]{Hacking1992}
Hacking I (1992) The self-vindication of the laboratory sciences. In: Pickering
  A (ed) Science as Practice and Culture, University of Chicago Press, pp
  29--64

\bibitem[{Hardwicke et~al.(2018)Hardwicke, Mathur, MacDonald, Nilsonne, Banks,
  Kidwell, Hofelich~Mohr, Clayton, Yoon, Henry~Tessler et~al.}]{Hardwicke2018}
Hardwicke TE, Mathur MB, MacDonald K, Nilsonne G, Banks GC, Kidwell MC,
  Hofelich~Mohr A, Clayton E, Yoon EJ, Henry~Tessler M, et~al. (2018) Data
  availability, reusability, and analytic reproducibility: Evaluating the
  impact of a mandatory open data policy at the journal cognition. Royal
  Society open science 5(8):180448

\bibitem[{Heesen(2018)}]{Heesen2018}
Heesen R (2018) Why the reward structure of science makes reproducibility
  problems inevitable. The Journal of Philosophy 115(12):661--674

\bibitem[{Ioannidis(2014)}]{Ioannidis2014}
Ioannidis JPA (2014) How to make more published research true. PLoS Medicine
  11(10):e1001747

\bibitem[{Iqbal et~al.(2016)Iqbal, Wallach, Khoury, Schully, and
  Ioannidis}]{Iqbal2016}
Iqbal SA, Wallach JD, Khoury MJ, Schully SD, Ioannidis JP (2016) Reproducible
  research practices and transparency across the biomedical literature. PLoS
  biology 14(1):e1002333

\bibitem[{Janssen et~al.(2012)Janssen, Charalabidis, and
  Zuiderwijk}]{Janssen2012}
Janssen M, Charalabidis Y, Zuiderwijk A (2012) Benefits, adoption barriers and
  myths of open data and open government. Information systems management
  29(4):258--268

\bibitem[{Kerr(1998)}]{Kerr1998}
Kerr NL (1998) Harking: Hypothesizing after the results are known. Personality
  and Social Psychology Review 2(3):196--217

\bibitem[{LeBel and Peters(2011)}]{Lebel2011}
LeBel EP, Peters KR (2011) Fearing the future of empirical psychology: Bem's
  (2011) evidence of psi as a case study of deficiencies in modal research
  practice. Review of General Psychology 15(4):371--379

\bibitem[{Leonelli(2018)}]{Leonelli2018}
Leonelli S (2018) Rethinking reproducibility as a criterion for research
  quality. In: Including a Symposium on Mary Morgan: Curiosity, Imagination,
  and Surprise, Emerald Publishing Limited, pp 129--146

\bibitem[{Loken and Gelman(2017)}]{Loken2017}
Loken E, Gelman A (2017) Measurement error and the replication crisis. Science
  355(6325):584--585, \doi{10.1126/science.aal3618}

\bibitem[{McElreath and Smaldino(2015)}]{McElreath2015}
McElreath R, Smaldino PE (2015) Replication, communication, and the population
  dynamics of scientific discovery. {PLOS} {ONE} 10(8):e0136088,
  \doi{10.1371/journal.pone.0136088}

\bibitem[{Molloy(2011)}]{Molloy2011}
Molloy JC (2011) The open knowledge foundation: open data means better science.
  PLoS biology 9(12):e1001195

\bibitem[{Munaf{\`o} et~al.(2017)Munaf{\`o}, Nosek, Bishop, Button, Chambers,
  du~Sert, Simonsohn, Wagenmakers, Ware, and Ioannidis}]{Munafo2017}
Munaf{\`o} MR, Nosek BA, Bishop DVM, Button KS, Chambers CD, du~Sert NP,
  Simonsohn U, Wagenmakers EJ, Ware JJ, Ioannidis JPA (2017) A manifesto for
  reproducible science. Nature Human Behaviour 1(1):0021,
  \doi{10.1038/s41562-016-0021}

\bibitem[{{National Academies of Sciences, Engineering, and
  Medicine}(2017)}]{national2017fostering}
{National Academies of Sciences, Engineering, and Medicine} (2017) Fostering
  integrity in research. National Academies Press, Washington, D.C.,
  \doi{10.17226/21896}

\bibitem[{Newton(1676)}]{Newton1676}
Newton I (1676) Mr. newton's answer to the precedent letter, sent to the
  publisher. Philosophical Transactions of the Royal Society 11:698--705

\bibitem[{Nosek et~al.(2015)Nosek, Alter, Banks, Borsboom, Bowman, Breckler,
  Buck, Chambers, Chin, Christensen et~al.}]{Nosek2015}
Nosek BA, Alter G, Banks GC, Borsboom D, Bowman SD, Breckler SJ, Buck S,
  Chambers CD, Chin G, Christensen G, et~al. (2015) Promoting an open research
  culture. Science 348(6242):1422--1425

\bibitem[{Nosek et~al.(2018)Nosek, Ebersole, DeHaven, and Mellor}]{Nosek2018}
Nosek BA, Ebersole CR, DeHaven AC, Mellor DT (2018) The preregistration
  revolution. Proceedings of the National Academy of Sciences
  115(11):2600--2606

\bibitem[{Penders et~al.(2019)Penders, Holbrook, and de~Rijcke}]{Penders2019}
Penders B, Holbrook JB, de~Rijcke S (2019) Rinse and repeat: Understanding the
  value of replication across different ways of knowing. Publications 7(3):52

\bibitem[{Popper(1959)}]{Popper1959}
Popper KR (1959) The logic of scientific discovery. University Press

\bibitem[{Quine(1976)}]{Quine1976}
Quine WvO (1976) Two dogmas of empiricism. In: Can Theories be Refuted?,
  Springer, pp 41--64, \doi{10.1007/978-94-010-1863-0\_2}

\bibitem[{Radder(1992)}]{Radder1992}
Radder H (1992) Experimental reproducibility and the experimenters' regress.
  In: PSA: Proceedings of the Biennial Meeting of the Philosophy of Science
  Association, Philosophy of Science Association, vol 1992, pp 63--73

\bibitem[{Radder(1996)}]{Radder1996}
Radder H (1996) In and about the world: Philosophical studies of science and
  technology. SUNY Press

\bibitem[{Ribe(1985)}]{Ribe1985}
Ribe NM (1985) Goethe's critique of {Newton}: A reconsideration. Studies in
  History and Philosophy of Science Part A 16(4):315--335,
  \doi{10.1016/0039-3681(85)90015-9}

\bibitem[{Schmidt(2009)}]{Schmidt2009}
Schmidt S (2009) Shall we really do it again? the powerful concept of
  replication is neglected in the social sciences. Review of General Psychology
  13(2):90--100

\bibitem[{Silberzahn et~al.(2018)Silberzahn, Uhlmann, Martin, Anselmi, Aust,
  Awtrey, Bahn{\'{\i}}k, Bai, Bannard, Bonnier, Carlsson, Cheung, Christensen,
  Clay, Craig, Rosa, Dam, Evans, Cervantes, Fong, Gamez-Djokic, Glenz,
  Gordon-McKeon, Heaton, Hederos, Heene, Mohr, Högden, Hui, Johannesson,
  Kalodimos, Kaszubowski, Kennedy, Lei, Lindsay, Liverani, Madan, Molden,
  Molleman, Morey, Mulder, Nijstad, Pope, Pope, Prenoveau, Rink, Robusto,
  Roderique, Sandberg, Schlüter, Schönbrodt, Sherman, Sommer, Sotak, Spain,
  Spörlein, Stafford, Stefanutti, Tauber, Ullrich, Vianello, Wagenmakers,
  Witkowiak, Yoon, and Nosek}]{Silberzahn2018}
Silberzahn R, Uhlmann EL, Martin DP, Anselmi P, Aust F, Awtrey E, Bahn{\'{\i}}k
  {\v{S}}, Bai F, Bannard C, Bonnier E, Carlsson R, Cheung F, Christensen G,
  Clay R, Craig MA, Rosa AD, Dam L, Evans MH, Cervantes IF, Fong N,
  Gamez-Djokic M, Glenz A, Gordon-McKeon S, Heaton TJ, Hederos K, Heene M, Mohr
  AJH, Högden F, Hui K, Johannesson M, Kalodimos J, Kaszubowski E, Kennedy DM,
  Lei R, Lindsay TA, Liverani S, Madan CR, Molden D, Molleman E, Morey RD,
  Mulder LB, Nijstad BR, Pope NG, Pope B, Prenoveau JM, Rink F, Robusto E,
  Roderique H, Sandberg A, Schlüter E, Schönbrodt FD, Sherman MF, Sommer SA,
  Sotak K, Spain S, Spörlein C, Stafford T, Stefanutti L, Tauber S, Ullrich J,
  Vianello M, Wagenmakers EJ, Witkowiak M, Yoon S, Nosek BA (2018) Many
  analysts, one data set: Making transparent how variations in analytic choices
  affect results. Advances in Methods and Practices in Psychological Science
  1(3):337--356, \doi{10.1177/2515245917747646}

\bibitem[{Spellman(2015)}]{Spellman2015}
Spellman BA (2015) A short (personal) future history of revolution 2.0

\bibitem[{Stanley and Spence(2014)}]{Stanley2014}
Stanley DJ, Spence JR (2014) Expectations for replications: Are yours
  realistic? Perspectives on Psychological Science 9(3):305--318,
  \doi{10.1177/1745691614528518}

\bibitem[{Stodden(2011)}]{Stodden2011}
Stodden V (2011) Trust your science? open your data and code. Amstat News
  July:21–22

\bibitem[{Taper and Lele(2011)}]{Taper2011}
Taper ML, Lele SR (2011) Evidence, evidence functions, and error probabilities.
  In: Philosophy of Statistics, vol~7, Elsevier, Oxford, pp 513--532,
  \doi{10.1016/b978-0-444-51862-0.50015-0}

\bibitem[{van't Veer and Giner-Sorolla(2016)}]{Veer2016}
van't Veer AE, Giner-Sorolla R (2016) Pre-registration in social psychology—a
  discussion and suggested template. Journal of Experimental Social Psychology
  67:2--12

\bibitem[{Wagenmakers et~al.(2012)Wagenmakers, Wetzels, Borsboom, van~der Maas,
  and Kievit}]{Wagenmakers2012}
Wagenmakers EJ, Wetzels R, Borsboom D, van~der Maas HL, Kievit RA (2012) An
  agenda for purely confirmatory research. Perspectives on Psychological
  Science 7(6):632--638

\bibitem[{Weisberg(2012)}]{weisberg2012simulation}
Weisberg M (2012) Simulation and similarity: Using models to understand the
  world. Oxford University Press

\bibitem[{Westfall and Devons(1981)}]{Westfall1981}
Westfall RS, Devons S (1981) Never at rest: A biography of {Isaac Newton}.
  Cambridge University Press, Cambridge, \doi{10.1017/cbo9781107340664}

\end{thebibliography}

%\begin{thebibliography}{}
% etc
%\end{thebibliography}

\end{document}